\documentclass[a4paper,12pt]{article}

\usepackage{graphicx}
\usepackage{array}
\usepackage{amsmath}
\usepackage{amssymb}
\usepackage{latexsym}
\usepackage{subfigure}
\newcommand{\vev}[1]{\langle{#1}\rangle}
\newcommand{\der}{\partial}
\newcommand{\dg}{\dagger}

\usepackage[normalem]{ulem}
\usepackage[dvips]{color}
\renewcommand\sout{\bgroup \color{red} \ULdepth=-.5ex \ULset}

\date{empty}
\begin{document}
\begin{titlepage}
\null
\begin{flushright}
June, 2017
\end{flushright}
\vskip 2.0cm
\begin{center}
{\Large \bf 
Spatially Modulated Vacua \\ 
in a Lorentz-invariant Scalar Field Theory
}
\vskip 1.7cm
\normalsize
\renewcommand\thefootnote{\alph{footnote}}

{\large
Muneto Nitta$^{\dagger}$\footnote{nitta(at)phys-h.keio.ac.jp},
Shin Sasaki$^\ddagger$\footnote{shin-s(at)kitasato-u.ac.jp}
and 
Ryo Yokokura$^\sharp$\footnote{ryokokur(at)keio.jp}
}
\vskip 0.7cm
{\it
$^\dagger$ 
Department of Physics, and Research and Education Center for Natural Sciences, \\
\vskip -0.2cm
Keio University, Hiyoshi 4-1-1, Yokohama, Kanagawa 223-8521, Japan  
\vskip 0.1cm
$^\ddagger$
Department of Physics,  Kitasato University, Sagamihara 252-0373, Japan
\vskip 0.1cm
$^\sharp$
Department of Physics, Keio University, Yokohama 223-8522, Japan
}
\vskip 0.5cm
\begin{abstract}
Spatial modulation has been studied for a long time in 
condensed matter, nuclear matter and quark matter, 
where the manifest Lorentz invariance is lost due to the finite density/temperature effects and so on.
In this paper, 
spatially modulated vacua at zero temperature and 
zero density are studied in Lorentz invariant field theories.
We first propose an adaptation
of the Nambu-Goldstone theorem 
to higher derivative theories
under the assumption of the absence of ghosts: 
when a global symmetry is spontaneously broken due to 
vacuum expectation values of 
space-time derivatives of fields, 
a Nambu-Goldstone (NG) boson appears 
without a canonical kinetic (quadratic derivative) term 
with a quartic derivative term in the modulated direction 
while a Higgs boson appears with a canonical kinetic term.
We demonstrate this in a simple model 
allowing (meta)stable modulated vacuum 
of a phase modulation (Fulde-Ferrell state),
where an NG mode associated with spontaneously broken 
 translational and $U(1)$ symmetries appears.

\end{abstract}
\end{center}

\end{titlepage}

\newpage

\setcounter{footnote}{0}
\renewcommand\thefootnote{\arabic{footnote}}
\pagenumbering{arabic}

\newpage
\section{Introduction}

Spatially modulated ground states were theoretically proposed 
 in superconductors 
a half century ago 
\cite{Fulde:1964zz,larkin:1964zz},
and such states are now called 
Fulde-Ferrell-Larkin-Ovchinnikov (FFLO) states. 
More precisely, Fulde-Ferrell (FF) and 
Larkin-Ovchinnikov (LO) states 
denote modulations of a phase and amplitude of 
a condensation, respectively. 
The LO states were shown to be ground states in  
the presence of a magnetic field inducing 
the spin imbalance for a Cooper pair of a superconductor 
\cite{Machida:1984zz}. 
In the last couple of years, 
there have been several claims of its observation 
(see Ref.~\cite{FFLO-review} for a review). 
Recently, ultracold atomic Fermi gases have renewed interest 
in FFLO states (see Ref.~\cite{Radzihovsky:2010} for a review).
The spin polarized superfluid state was observed in Ref.~\cite{Liao:2010} 
 and it was claimed that the 
FFLO state has been achieved in this experiment.
FFLO states in a ring were also 
proposed in cold Fermi gases \cite{Yanase:2009} 
and in superconductors \cite{Yoshii:2014fwa}.

FFLO states, 
called twisted kink crystals, 
were also studied in the 
chiral Gross-Neveu model in 1+1 dimensions 
\cite{Basar:2008im,Basar:2008ki,Basar:2009fg}
(see \cite{Yoshii:2011yt} for application to a superconductor).
Spatially modulated chiral condensations,  
such as FF states (called dual chiral density wave or chiral spiral)
\cite{Nakano:2004cd,Karasawa:2013zsa} 
and LO states (called real kink crystal)
\cite{Nickel:2009ke,Nickel:2009wj}, 
have been proposed  to appear in a certain region 
of the phase diagram of QCD 
in 3+1 dimensions (see Ref.~\cite{Buballa:2014tba} as a review).
%\Blue{
Although the Cooper pair is usually refers to the particle-particle condensates, 
the chiral condensation is related to the particle-antiparticle (or hole) pairing.
%}
They were also proposed in diquark condensations 
exhibiting 
color superconductivity in high density QCD 
(see \cite{Casalbuoni:2003wh,Anglani:2013gfu} as a review) 
and were also discussed 
in the context of the AdS/CFT correspondence 
\cite{Nakamura:2009tf, Amoretti:2016bxs, Andrade:2017leb, Cai:2017qdz}.

These spatial modulations 
were originally proposed in condensations of fermions 
forming Cooper pairs. 
In terms of the Ginzburg-Landau effective theory, 
which is a scalar field theory,
these states are 
realized
as ground states of the theory 
due to the presence of a wrong sign of a gradient term
and positive higher derivative terms. 
In general, these kinds of inhomogeneous states
spontaneously break translational 
as well as rotational symmetries.
Nambu-Goldstone (NG) modes 
associated with these broken symmetries
in such backgrounds
were studied in Refs.~\cite{Lee:2015bva,Hidaka:2015xza}. 
After all, inhomogeneous states 
in condensed matter, nuclear matter 
and quark matter
studied so far are all realized in theories where the Lorentz invariance is explicitly 
broken due to the finite density/temperature effects and so on.

In this paper, we study spatially modulated vacua at 
zero temperature and zero density  
(but not ground states in finite density and/or temperature)
in manifestly Lorentz invariant field theories, 
with a particular attention to spontaneous symmetry breaking 
and NG bosons. 
From a viewpoint of low-energy effective theories, 
field theories generically receive higher derivative corrections. 

We assume that there is no ghost in the theory implying the absence of 
more than one derivative on one field which can not be eliminated by partial integration.
For example, the term 
$\partial^2 \varphi = \partial_m \partial^m \varphi $ 
with space-time index $m$, 
generally causes the so-called Ostrogradski instability \cite{Ostrogradski}. 
This is a crucial difference with non-relativistic cases.  
Then, all higher derivative terms come in a way 
that only a single space-time derivative
acting on one field, $\partial^m \varphi$. 
Thus, the effective theory is in general 
a function of $\partial^m \varphi$ (complemented by a potential term).
In this set-up 
 we study 
an adaptation of the NG theorem to higher derivative theories
, stating that when a global symmetry is spontaneously broken due to vacuum expectation values of 
space-time derivatives of fields, 
an NG boson appears 
without canonical kinetic (quadratic derivative) terms 
with a quartic derivative term in the modulated direction,
while a Higgs boson appears 
with a non-zero canonical kinetic term.

After giving general discussion of the stability of 
general higher derivative models,
we give a simple model illustrating this.
Our model admits (meta)stable modulated vacuum 
of a phase modulation (Fulde-Ferrell state),
where an NG mode associated with spontaneously broken 
 translational and $U(1)$ symmetries appears.

\section{
Adaptation of the Nambu-Goldstone theorem to higher derivative theories
}
In this section, we apply 
the NG theorem to the case that 
global symmetries of a Lagrangian are 
spontaneously broken due to 
vacuum expectation values (VEVs) of 
space-time derivatives of fields. 
Here, we consider the case that there is no 
Ostrogradski instability 
\cite{Ostrogradski} 
assuming that there is no more 
than one space-time derivative on a field.
We show that
an analogue of the NG boson appears 
without canonical kinetic term
with a quartic derivative term. 
In addition, we will show that a 
Higgs boson, which is defined by a mode that is orthogonal to the
abovementioned NG mode, appears with a non-zero canonical kinetic term
in the vacuum.

In the following, we consider 
$d$-dimensional
relativistic field theories where the Lorentz invariant Lagrangian
$\mathcal{L}$ is given by a functional of $\partial_m
\varphi_a$. 
Here $m = 0,1, \ldots, d-1$ is the space-time index and 
$\varphi_a \ (a=1, \ldots, N)$ are complex scalar fields.
The energy functional $\mathcal{E}$ of the
theories depends only on the first time derivative of fields which we
denote $\Phi_I = \partial_m \varphi_a, \Phi_I^{\dagger} = 
\partial_m \bar{\varphi}_a$. The index $I = 1, \ldots, dN$ labels the
fields and directions of the space-time derivatives.
Vacua $|0 \rangle$ in the theories are defined such a way that they provide extrema of the energy
$\mathcal{E}$ with respect to $\Phi_I, \Phi_I^{\dagger}$:
\begin{align}
\frac{\partial \mathcal{E}}{\partial \Phi_I} = \frac{\partial
 \mathcal{E}}{\partial \Phi^{\dagger}_I} = 0.
\label{eq:extrema}
\end{align}
In these extrema, we assume that the fields $\Phi_I, \Phi^{\dagger}_I$
develop VEVs:
\begin{align}
\langle 0 | \Phi_I | 0 \rangle = v_I, \qquad 
\langle 0 | \Phi_I^{\dagger} | 0 \rangle = \bar{v}_I
\label{eq:vevs}.
\end{align}
Here some of these VEVs are not zero and they need not to be constants
in general.
Indeed, as we will see later, they are spatially varying functions for
modulated vacua.
Since $\Phi_I$ are in fact given by the space-time derivative of the fields
(therefore they are Lorentz vectors), the non-zero VEVs \eqref{eq:vevs} generically
break the translational and rotational symmetries.
Hereafter, we assume that VEVs are spacelike vectors
\footnote{
The Higgs mechanism caused by non-zero VEVs of the time derivative of scalar
fields is known as the ghost condensation \cite{ArkaniHamed:2003uy}.
Henceforth, we never consider VEVs of timelike vectors.
}.

Now we introduce the dynamical fields 
$\tilde{\Phi}_I = \partial_m \tilde{\varphi}_a, \tilde{\Phi}^{\dagger}_I
= \partial_m \tilde{\varphi}^{\dagger}_a $
as fluctuations around a vacuum determined by 
the condition
\eqref{eq:extrema}.
We shift the fields around the VEVs, $\Phi_I \to v_I + \tilde{\Phi}_I$, 
and the energy is expanded as
\begin{align}
\mathcal{E} (v + \tilde{\Phi}, \bar{v} + \tilde{\Phi}^{\dagger}) =
 \mathcal{E} (v,\bar{v}) + \frac{1}{2} 
(\tilde{\Phi}^{\dagger}_I, \tilde{\Phi}_I)
M_{IJ}
\left(
\begin{array}{c}
\tilde{\Phi}_J \\
\tilde{\Phi}_J^{\dagger}
\end{array}
\right)
+ \cdots,
\label{eq:energy_general}
\end{align}
where we have defined the following Hermitian matrix:
\begin{align}
M_{IJ} = \left(
\begin{array}{cc}
\left. \frac{\partial^2 \mathcal{E}}{\partial \Phi^{\dagger}_I \partial
 \Phi_J} \right|_v 
& 
\left. \frac{\partial^2 \mathcal{E}}{\partial \Phi^{\dagger}_I \partial
 \Phi^{\dagger}_J} \right|_v 
\\
\left. \frac{\partial^2 \mathcal{E}}{\partial \Phi_I \partial
 \Phi_J} \right|_v
&
\left. \frac{\partial^2 \mathcal{E}}{\partial \Phi_I \partial
 \Phi^{\dagger}_J} \right|_v
\end{array}
\right).
\label{eq:matrix}
\end{align}
Here the symbol $ * |_v$ stands for the values evaluated in the vacuum.
We note that the matrix $M$ determined by the second derivatives of
$\mathcal{E}$ is just the curvature of the energy density and it is 
in general a function of $x^i \ (i=1,2,3)$.
In order that the extrema defined by \eqref{eq:extrema} become local
minima of the energy, $M$ should be a positive semi-definite matrix
for all the regions in $x$.

These conditions of vacua do not guarantee that they are global minima but 
meta-stable local minima are allowed in general.
From the expression \eqref{eq:energy_general}, one observes that the eigenvalues
of $M$ correspond to coefficients of the quadratic kinetic terms of 
the dynamical fields $\tilde{\varphi}_a, \tilde{\varphi}^{\dagger}_a$.
Since $M$ is a positive semi-definite matrix, there are no fluctuation modes whose 
 kinetic terms have the wrong sign 
(negative sign in the energy functional).
However we stress that there are possible zero eigenvalues for a general $M$.
When $M$ has zero eigenvalues, the quadratic terms of the corresponding modes vanish.

In order to see the meaning of the zero eigenvalues of $M$, 
we elucidate the relation between the matrix $M$ and the spontaneous
symmetry breaking.
The fields $\Phi_I, \Phi^{\dagger}_I$ transform according to symmetries
of theories:
\begin{align}
[i \mathcal{Q}^A, \Phi_I] = i (T^A \Phi)_I, \qquad 
[i \mathcal{Q}^A, \Phi^{\dagger}_I] = - i (T^A \Phi)^{\dagger}_I.
\end{align}
Here $\mathcal{Q}^A$ are generators of the symmetry groups and 
the Hermitian matrices $T^A$ are an irreducible representation of $\mathcal{Q}^A$.
In a vacuum $|0 \rangle$, some of the fields $\Phi_I$ develop non-zero VEVs and we have 
\begin{align}
\langle 0 | [i \mathcal{Q}^A, \Phi_I] | 0 \rangle =  i (T^A v)_I, \qquad 
\langle 0 | [i \mathcal{Q}^A, \Phi^{\dagger}_I] | 0 \rangle = - i
(T^A v)^{\dagger}_I.
\end{align}
We now define the following vector,
\begin{align}
(T^A \vec{v})_I \equiv 
\left(
\begin{array}{c}
(T^A v)_I \\
- (T^A v)^{\dagger}_I
\end{array}
\right).
\end{align}
For generators that satisfy $T^{\hat{A}} \vec{v} = 0$, the corresponding symmetry is
preserved in the vacuum while for $T^{A'} \vec{v} \not= 0$ the symmetry is spontaneously broken.
The energy functional $\mathcal{E}$ is invariant under the following transformation,
\begin{align}
\Phi_I \to \Phi_I + i \varepsilon^A (T^A \Phi)_I, \qquad 
\Phi_I^{\dagger} \to \Phi_I^{\dagger} - i \varepsilon^A (T^A \Phi)^{\dagger}_I,
\end{align}
where $\varepsilon^A$ are infinitesimal real parameters.
Then we have
\begin{align}
\frac{\partial \mathcal{E}}{\partial \Phi_J} (T^A \Phi)_J -
 \frac{\partial \mathcal{E}}{\partial \Phi^{\dagger}_J} (T^A \Phi)^{\dagger}_J
 = 0.
\end{align}
By differentiating the above relation with respect to $\Phi_I, \Phi^{\dagger}_I$ and
evaluate the result in a vacuum, we find
\begin{align}
 M_{IJ} (T^{A'} \vec{v})_J = 0.
\label{eq:eigen}
 \end{align}
Therefore $T^{A'} \vec{v}$ are eigenvectors associated with the zero eigenvalues of $M$.
The relation \eqref{eq:eigen} indicates that
when some of the fields
$\Phi_I, \Phi_I^{\dagger}$ develop VEVs that spontaneously break
symmetries, then the canonical quadratic kinetic terms for the modes that
correspond to the zero eigenvalues of $M$ vanish. We call these 
Nambu-Goldstone (NG) modes.
On the other hand, the modes that are orthogonal to the NG modes appear
with quadratic kinetic terms in the energy functional. We call these Higgs
modes. 
We note that since the vector $T^{A'} \vec{v}$ generically depends on
$x^i$, there is a possibility that $T^{A'} \vec{v} (x)$ vanishes at some
specific points $x^i = x^i_0$ in a general setup.
At these points, the broken symmetries are recovered
locally and one expects that a non-zero quadratic term associated with the NG mode recovers.
We do not exclude this possibility but it is nevertheless not always the case.

Indeed, as we will show in an explicit example of the spatially
dependent VEV in the next section, 
the vector $T^{A'} \vec{v}$ never vanishes at special points and the
theorem discussed in this section completely works well in all regions in space-time.

\section{A model for spatially modulated vacua}
In order to understand the discussion in the previous section concretely, 
we introduce a Lorentz invariant scalar field model where, in addition to
the canonical quadratic kinetic terms, higher derivative corrections are
involved.
We begin with the observation that the global stability of modulation is
guaranteed when the highest power of the derivative terms $|\partial_m
\varphi|^2$ is odd and an appropriate sign of the terms are chosen.
We propose a scalar field model of the simplest example where a spatially
modulated state is allowed as a (meta-)stable vacuum.
We then apply the Nambu-Goldstone theorem discussed in the previous
section to the model and show that there are modes where
the quadratic kinetic terms vanish (NG modes).
We demonstrate that there are always associated modes with the non-zero quadratic kinetic
terms (Higgs modes).

\subsection{Global stability of modulation}
Let us consider a complex scalar field $\varphi$. 
The general Lorentz invariant Lagrangian containing finite powers of 
$|\partial \varphi|^2 = \partial_m \varphi \partial^m
\bar{\varphi}$ is 
\begin{align}
 {\cal L} 
  = \mp |\partial \varphi|^{2n} + \cdots
  = \mp \left| - |\dot \varphi|^2 + |\nabla \varphi|^2 \right|^n
 +\cdots,
\label{eq:Higher_derivative_Lagrangian}
\end{align}
where $n \in \mathbb{Z}$ is the highest power of the derivative terms
and $\cdots$ implies lower orders. 
The space-time index $m$
is contracted by $\eta_{mn}=\text{diag.}(-1,1,1,1)$.
The dot in $\dot{\varphi}$ stands for the derivative of the field with respect to $x^0$
and $\nabla$ is spatial derivatives.
The canonical conjugate momentum is
\begin{align}
 \pi_{\varphi} = {\partial {\cal L} \over \partial \dot\varphi} 
 = \mp (-1)^n n \dot{\bar \varphi}  |\partial \varphi|^{2n-2} + \cdots.
\end{align}
Then, the Hamiltonian associated with the Lagrangian
\eqref{eq:Higher_derivative_Lagrangian} reads
\begin{align}
 \mathcal{H} = & \ \pi_{\varphi} \dot{\varphi} + \pi_{\bar{\varphi}}
 \dot{\bar{\varphi}} - \mathcal{L}  
 =  \mp (-1)^n (2n-1) |\partial \varphi|^{2n} 
 \pm  |\nabla \varphi|^{2n} \cdots.
\label{eq:Higher_derivative_Hamiltonian}
\end{align}

Let us discuss the stability 
of a vacuum in the model.
First, by looking at the second term in \eqref{eq:Higher_derivative_Hamiltonian},
we see that the energy is bounded from below only when one chooses 
the upper sign in \eqref{eq:Higher_derivative_Lagrangian}. 
Otherwise, 
the spatial gradient of the field causes an instability as 
$|\nabla \varphi|^2 \to \infty$.
Second, the first term in \eqref{eq:Higher_derivative_Hamiltonian}
implies that the energy is bounded from below only when the highest order 
$n$ is odd. For even $n$, an instability in the temporal direction grows 
$|\dot \varphi|^2 \to \infty$.
Therefore, the simplest Lagrangian is of the third order in 
$|\partial \varphi|^2$ containing six derivatives.
In the next subsection, we consider 
an example of the third order Lagrangian allowing 
a modulated vacuum. 

%%%%%%%%%%%%%%%%%%
\subsection{A model and vacua}
We propose a four-dimensional Lorentz invariant complex scalar field model whose
Lagrangian is given by 
\begin{align}
\mathcal{L} =& \ 
- k \partial_m \varphi \partial^m \bar{\varphi}
+ 
(\lambda - \alpha \partial_m \varphi \partial^m \bar{\varphi})
(\partial_n \varphi \partial^n \varphi) (\partial_p \bar{\varphi}
 \partial^p \bar{\varphi}).
\label{eq:Lagrangian}
\end{align}
Here $k> 0, \lambda > 0, \alpha > 0 
$ are real constants
\footnote{
In a realistic setup, for example, such as Gross-Neveu models, low-energy effective theory of high density QCD,
these parameters are determined by the dynamics of the UV regime. 
See for example \cite{Nickel:2009ke, Abuki:2011pf, Carignano:2017meb}.
}.
The Lagrangian \eqref{eq:Lagrangian} 
contains 
 the ordinary kinetic term of the complex scalar field $\varphi$ and the higher derivative corrections.
The Lagrangian \eqref{eq:Lagrangian} 
is invariant under a global $U(1)$
transformation $\varphi \to e^{i \theta} \varphi$
with a constant $\theta$,
in addition to the Poincar\'e symmetry including the 
$SO(3,1)$ Lorentz and translational symmetries.
The Lagrangian can contain a potential term of $\varphi$ too. 
In this paper, for simplicity we do not consider a potential term.
In this case, the Lagrangian possesses a shift symmetry 
\begin{align}
\varphi \to \varphi +c,
\label{eq:shift}
\end{align}
where $c$ is a constant.

The energy functional associated with the Lagrangian
\eqref{eq:Lagrangian} is calculated to be 
\begin{align}
\mathcal{E} =& \ \pi_{\varphi} \dot{\varphi} + \pi_{\bar{\varphi}}
 \dot{\bar{\varphi}} - \mathcal{L} 
\notag \\
=& \ k (|\dot{\varphi}|^2 + |\partial_i
 \varphi|^2) + \left\{\lambda - \alpha (- |\dot{\varphi}|^2 +
 |\partial_i \varphi|^2) \right\} 
\left\{
3 |\dot{\varphi}|^4 - \dot{\varphi}^2 (\partial_i \bar{\varphi})^2 -
 \dot{\bar{\varphi}}^2 (\partial_i \varphi)^2 - (\partial_i \varphi)^2
 (\partial_j \bar{\varphi})^2
\right\}
\notag \\
+& \  
2 \alpha |\dot{\varphi}|^2
\left\{
|\dot{\varphi}|^4 - \dot{\varphi}^2 (\partial_i \bar{\varphi})^2 -
 \dot{\bar{\varphi}}^2 (\partial_i \varphi)^2 + (\partial_i \varphi)^2
 (\partial_j \bar{\varphi})^2
\right\},
\label{eq:energy}
\end{align}
where $i,j = 1,2,3$, $\pi_{\varphi}, \pi_{\bar{\varphi}}$ are the canonical momenta for
$\varphi$, $\bar{\varphi}$ and $\dot{\varphi} =
\frac{\partial\varphi}{\partial x^0}, \dot{\bar{\varphi}} =
\frac{\partial \bar{\varphi}}{\partial x^0}$.
The spatially modulated vacua (ground states) are expected to appear at extrema of
the energy functional \eqref{eq:energy}.

\begin{figure}[t]
\begin{center}
\subfigure[$\mathcal{E} (x_+) > 0$ ($\alpha = 1, \lambda = 3.8, k = 4$)]
{
\includegraphics[scale=.57]{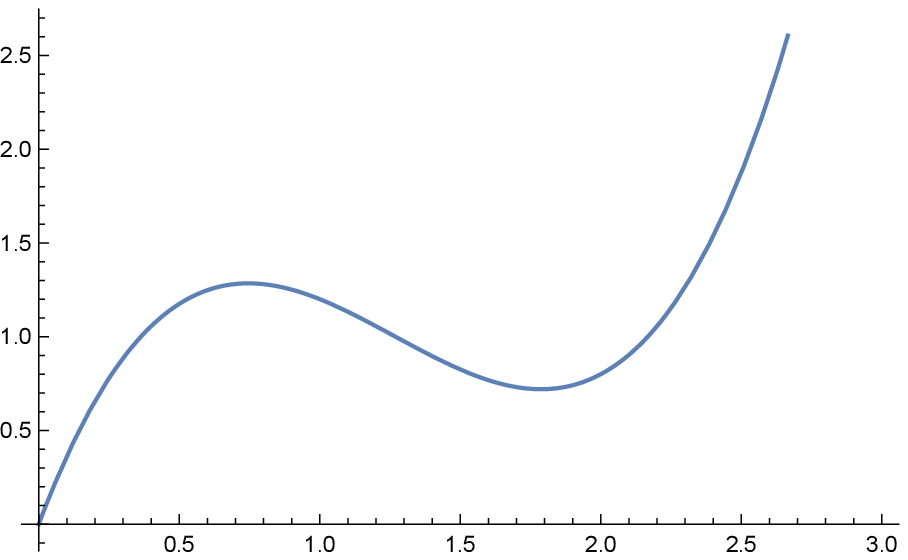}
}
\subfigure[$\mathcal{E} (x_+) = 0$ ($\alpha = 1,\lambda = 2, k = 1$)]
{
\includegraphics[scale=.57]{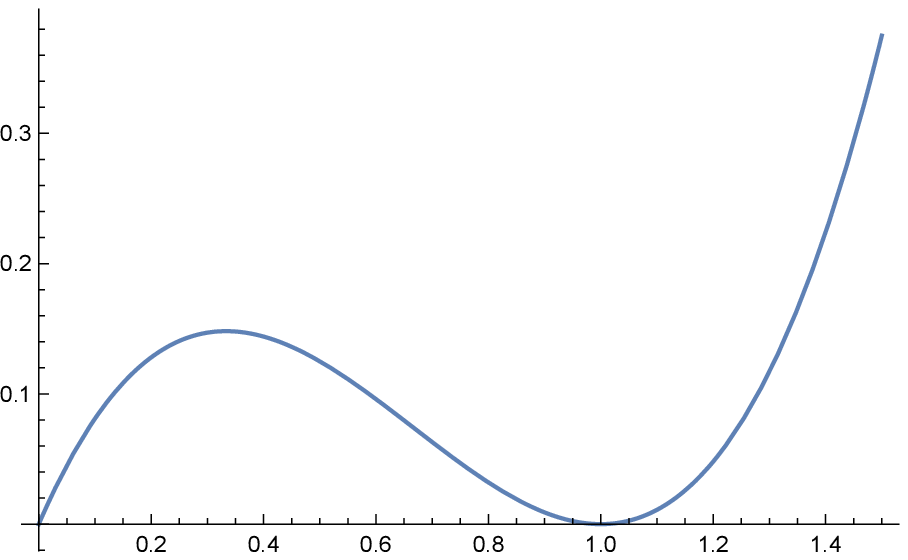}
}
\subfigure[$\mathcal{E} (x_+) < 0$ ($\alpha = 1, \lambda = 3, k = 1$)]
{
\includegraphics[scale=.57]{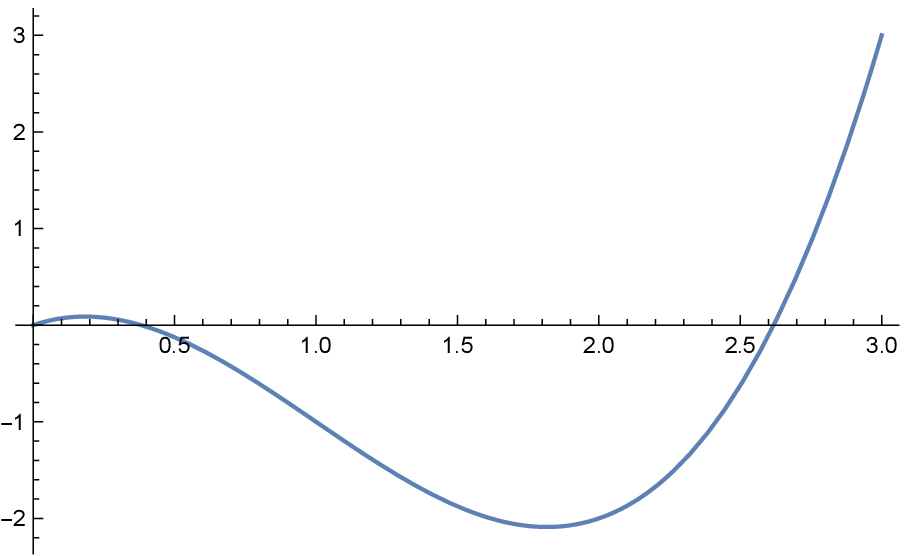}
}
\end{center}
\caption{Schematic pictures of the energy function $\mathcal{E} (x)$,
 the potential for $x = |\partial_1 \varphi|^2$. 
The vertical axis stands for the energy $\mathcal{E} (x)$ while the
 horizontal axis for $x$.
The local vacua for (a) positive, (b) zero and (c) negative vacuum
 energies are shown. Examples of the parameters that satisfy the
 conditions in the main body of the text are shown.
}
\label{fig:energy_functional}
\end{figure}

We now employ an ansatz
$\varphi = \varphi (x^1)$ for one-dimensional spatial modulation along
the $x^1$-direction: 
$\langle 0| \partial_1 \varphi | 0 \rangle \neq 0$. We also assume static configurations.
Then the energy functional
\eqref{eq:energy} becomes 
\begin{align}
\mathcal{E} =& \ 
\alpha |\partial_1 \varphi|^6
- \lambda |\partial_1 \varphi|^4 
+ k |\partial_1 \varphi|^2.
\end{align}
The function $\mathcal{E} (x)$ is interpreted as a potential for $x =
|\partial_1 \varphi|^2 \ge 0$.
It is obvious that $\mathcal{E} (x) (x \ge 0)$ has a local minimum at
$x = 0$ in which the vacuum energy is $\mathcal{E} (0) = 0$ and the
scalar field $\varphi$ has a constant VEV.
Whether $\mathcal{E} (x)$ has another minimum or not crucially depends
on the parameters $k, \lambda, \alpha$.
Since $\mathcal{E}'(x) = 3 \alpha x^2 - 2 \lambda x + k$, 
if the condition $\lambda^2 - 3 \alpha k > 0$ is satisfied, $\mathcal{E}
(x)$ has another vacuum. In this case, the function $\mathcal{E}(x)$ has extrema at
\begin{align}
x_{\pm} =  
\frac{\lambda \pm \sqrt{\lambda^2 - 3 \alpha k}}{3 \alpha}.
\label{eq:xplus}
\end{align}
Since $k > 0$, $x= x_{-}$ corresponds to a local maximum while 
$x = x_+ \not= 0$ is a minimum which is a candidate of a modulated vacuum.
Note that $\lambda$ should be positive in order that $x_{+} > 0$. 
The condition $\alpha >0$ is necessary in order that the potential is bounded from
 below.
We find that
 the vacuum energy is classified according to the discriminant condition of the function $\alpha x^2
- \lambda x + k$. We have three distinct types of vacua. 
When the parameters $k, \lambda, \alpha$ satisfy the 
condition $\lambda^2 - 4 \alpha k < 0$, then the function $\mathcal{E} (x)$ becomes positive definite.
The local vacuum energy at $x = x_+$ is positive $\mathcal{E} (x_+) > 0$.
This is a meta-stable vacuum which decays to the global minimum 
(true vacuum) $x = 0$
 within a finite time. See fig.\ref{fig:energy_functional} (a) for the
 potential profile.
When the parameters satisfy the condition $\lambda^2 - 4 \alpha k = 0$,
the potential function $\mathcal{E}(x)$ looks like Fig.~\ref{fig:energy_functional} (b).
In addition to the global vacuum $x = 0$, we have a local vacuum $x = x_+$
in which $\mathcal{E} (x_+) = 0$. They are actually global vacua and are degenerated.
Finally, when the condition $\lambda^2 - 4 \alpha k > 0$ is satisfied,
the function $\mathcal{E}(x)$ looks like fig.\ref{fig:energy_functional} (c).
Now the vacuum $x = 0$ becomes meta-stable and the vacuum at
 $x = x_{+}$
is energetically favoured. Then the true vacuum is located at $x = x_+$
in which $\mathcal{E}(x_+) < 0$.

In each vacuum we have $|\partial_1 \varphi|^2 = x_+$.
The general solution that satisfies this relation is 
\begin{align}
\varphi (x^1) = \sqrt{x_+} \int^{x_1}_c \! ds \ e^{i F(s)},
\label{eq:general_sol}
\end{align}
where $c$ is a constant and $F(s)$ is a real function.
We are interested in a spatially modulated vacuum state.
The most conservative choice is a linear function $F (s) = p s$ where
$p$ is a constant. As we will see below, 
this vacuum preserves the highest symmetry in the theory.
The vacuum is then given by 
\begin{align}
\varphi (x^1) = \varphi_0 e^{i p x^1}, \qquad p \in \mathbb{R}, \
 \varphi_0 \in \mathbb{C},
\label{eq:mod_vac}
\end{align}
where the constants $p, \varphi_0$ satisfy $p^2 |\varphi_0|^2 = x_{+}$. 
This is the ground state where the spatial modulation along the $x^1$-direction occurs.
The period of the modulation is given by $2 \pi/p$.
It is easy to confirm that the modulated vacuum \eqref{eq:mod_vac} satisfies
the equation of motion as follows.
The equation of motion for $\bar{\varphi}$ in the model
\eqref{eq:Lagrangian} is given by 
\begin{align}
\partial_m 
\left[
\frac{}{}
k \partial^m \varphi + \alpha \partial^m \varphi (\partial_n \varphi)^2
 (\partial_p \bar{\varphi})^2 - 2 (\lambda - \alpha \partial_n \varphi
 \partial^n \bar{\varphi}) (\partial_p \varphi)^2 \partial^m \bar{\varphi}
\right] = 0.
\end{align}
For a configuration $\varphi = \varphi (x^1)$, we have
\begin{align}
k \partial_1^2 \varphi + \partial_1 
\left[
\frac{}{}
\alpha \partial_1 \varphi |\partial_1 \varphi|^4 
- 2 (\lambda - \alpha |\partial_1 \varphi|^2) (\partial_1 \varphi)^2
 \partial_1 \bar{\varphi}
\right] = 0.
\label{eq:eom_modulation}
\end{align}
In the modulated vacuum \eqref{eq:mod_vac}, we have $|\partial_1 \varphi|^2 = x_+ =
\text{const}$. Therefore the equation \eqref{eq:eom_modulation} becomes
\begin{align}
(\partial_1^2 \varphi) 
\left[
3 \alpha x_+^2 - 2 \lambda x_+ + k
\right] = 0.
\end{align}
Since $3 \alpha x_+^2 - 2 \lambda x_+ + k = 0$, we have shown that the modulated vacuum
\eqref{eq:mod_vac} satisfies the equation of motion.

We have a spatially modulated vacuum \eqref{eq:mod_vac} along the $x^1$-direction
in which $\langle 0 | \partial_1 \varphi | 0 \rangle = v = i p \varphi_0 e^{ip x^1} \not= 0$.
It is obvious that 
the translational symmetry along the $x^1$-direction and the rotational symmetries in the $(x^1,x^2)$ and
$(x^1,x^3)$ planes are spontaneously broken in the modulated vacuum.
Then the four-dimensional Poincar\'e symmetry is broken down to that in
three dimensions.

As a matter of fact, due to the $U(1)$ symmetry $\varphi \to e^{i \theta } \varphi$, 
the simultaneous operation of the translation $x^1 \to x^1 + a$
and the $U(1)$ transformation $\varphi \to e^{- i p a} \varphi$ is preserved in the
modulated vacuum \eqref{eq:mod_vac}. 
Here $a$ is a constant.
Meanwhile, the combination of the translation and the inverse rotation
$x^1 \to x^1 + a$, $\varphi \to e^{+ i p a}\varphi$ is broken.
Therefore the global symmetry including the translational operation
along the $x^1$-direction is broken in such a way that $P^1 \times U(1) \to [P^1
\times U(1)]_{\text{sim}}$. Here $P^1$ represents translation along
the $x^1$-direction and ``sim'' means the simultaneous operation.
As we have remarked above, this symmetry breaking pattern is a
consequence of the simplest choice of the modulated vacuum 
\eqref{eq:mod_vac}. No other choices of $F(s)$ results in this $[P^1
\times U(1)]_{\text{sim}}$ symmetry.
Note that the translation $P^1$ and the rotations in the $(x^1,x^2)$ and
$(x^1,x^3)$ planes are not independent each other \cite{Low:2001bw}.
Therefore we expect that there is one NG mode associated 
with the spontaneous symmetry breaking $P^1 \times U(1) \to [P^1\times
U(1)]_{\text{sim}}$
\footnote{
Applications of this $P^1 \times U(1)$ symmetry to the pion condensate can be found, 
for example, in Refs.~\cite{Dautry:1979bk, Kutschera:1989yz}.
}
. 
We will clarify this issue in the followings subsections.

\subsection{Linear analysis: Nambu-Goldstone and Higgs bosons}
In this subsection, 
we show the stability analysis in our model, 
but  the  analysis employed here 
is completely general at the linear level 
for any model admitting local vacuum exhibiting 
a spatial modulation.
We now shift the field from the modulated vacuum \eqref{eq:mod_vac} and
introduce the fluctuation $\tilde{\varphi}$ as a dynamical field:
\begin{align}
\varphi \ \longrightarrow \ \langle \varphi \rangle + \tilde{\varphi},
\label{eq:fluct.}
\end{align}
where $\langle \varphi \rangle = \varphi_0 e^{i p x^1}$ is the
modulating VEV.
In the following, we will show that there are no fluctuation modes that cause 
instabilities of the vacuum \eqref{eq:mod_vac}.
The quadratic terms of the dynamical scalar field $\tilde{\varphi}$ are
extracted from the energy:
\begin{align}
\mathcal{E}_{\text{quad.}} =& \ 
 (k + \alpha x_+^2) \partial_{\hat{m}} \tilde{\varphi}
 \partial^{\hat{m}} \tilde{\varphi}^{\dagger} 
+ (\lambda - \alpha x_+) p^2 \varphi_0^2 e^{2 i p x^1}
 (\partial_{\hat{m}} \tilde{\varphi}^{\dagger})^2
+ (\lambda - \alpha x_+) p^2 \bar{\varphi}_0^2 e^{-2 i p x^1}
 (\partial_{\hat{m}} \tilde{\varphi})^2 
\notag \\
& \ 
+ (9 \alpha x_+^2 - 4 \lambda x_+ + k) \partial_1 \tilde{\varphi}
 \partial_1 \tilde{\varphi}^{\dagger}
+ (\lambda - 3 \alpha x_+) p^2 \varphi_0^2 e^{2 i p x^1} (\partial_1
 \tilde{\varphi}^{\dagger})^2 \notag\\
& \ + (\lambda - 3 \alpha x_+) p^2 \bar{\varphi}_0^2 e^{- 2 i p x^1}
	 (\partial_1 \tilde{\varphi})^2
\notag \\
=& \  \frac{1}{2} \vec{\varphi}^{\dagger} \mathbf{M}
 \vec{\varphi}.
\end{align}
Here the index $\hat{m} = 0,2,3$ is contracted by $\delta_{\hat{m} \hat{n}}$.
The vector $\vec{\varphi}$ and the matrix 
$\mathbf{M}$ are defined by
\begin{align}
& 
\vec{\varphi} =
\left(
\begin{array}{c}
\partial^{\hat{m}} \tilde{\varphi} \\
\partial^{\hat{m}} \tilde{\varphi}^{\dagger} \\
\partial_1 \tilde{\varphi} \\
\partial_1 \tilde{\varphi}^{\dagger}
\end{array}
\right), 
\qquad 
\mathbf{M} = 
\left(
\begin{array}{cc}
M_1 & 0 \\
0 & M_2
\end{array}
\right).
\label{eq:vector_matrix}
\end{align}
Each block diagonal sector is found to be
\begin{align}
M_1 =
\left(
\begin{array}{cc}
k + \alpha x_+^2 & 2 (\lambda -  \alpha x_+) p^2 \varphi_0^2 e^{2 i p
 x^1} \\
2 (\lambda - \alpha x_+) p^2 \bar{\varphi}_0^2 e^{-2 i p x^1} & k +
 \alpha x_+^2
\end{array}
\right), \notag \\
M_2  = 
\left(
\begin{array}{cc}
9 \alpha x_+^2 - 4 \lambda x_+ + k & 2 (\lambda - 3 \alpha x_+) p^2
 \varphi_0^2 e^{2 i p x^1} \\
2 (\lambda - 3 \alpha x_+) p^2 \bar{\varphi}_0^2 e^{-2 i p x^1} & 9
 \alpha x_+^2 - 4 \lambda x_+ + k
\end{array}
\right).
\end{align}
We have separated the quadratic terms to the $SO(2,1)$ Lorentz invariant
sector (transverse direction) and the direction of the modulation. 
Since $\mathbf{M}$ is an Hermitian matrix, it is diagonalized by a unitary matrix:
\begin{align}
\mathbf{U} = 
\left(
\begin{array}{cc}
U_1 & 0 \\
0 & U_2
\end{array}
\right).
\end{align}
Here $U_1, U_2$ are $2 \times 2$ unitary matrices.
The eigenvalues $s_1, s_2$ of $M_1$ are found to be 
\begin{align}
s_1 = 4 x_+ (\lambda - \alpha x_+),
\qquad s_2 = 3 \alpha x_{+}^2 - 2 \lambda x_+ + k.
\end{align}
It is easy to show that $s_1 > 0$ and $s_2 = 0$ 
by the definition of $x_+$.
The eigenvalues $t_1, t_2$ of $M_2$ are calculated as 
\begin{align}
t_1 = 4 x_+ (3 \alpha x_+ - \lambda), 
\qquad 
t_2 = 3 \alpha x_+^2 - 2 \lambda x_+ + k.
\end{align}
Again we find $t_1 > 0$ and $t_2 = 0$.
There are positive and zero eigenvalues as anticipated.
This implies that our assumption $\dot{\varphi} = \partial_2 \varphi =
\partial_3 \varphi = 0$ guarantees the minimization condition of the energy.
Therefore, the eigenvalues and eigenvectors of $\mathbf{M}$ are given by 
\begin{align}
&
s_1 > 0 \ : \ \mathbf{e}_1 = \frac{1}{\sqrt{2} |\varphi_0|} 
\left(
\begin{array}{c}
\langle \varphi \rangle \\
\langle \bar{\varphi} \rangle \\
0 \\
0
\end{array}
\right), \qquad 
s_2 = 0 \ : \ \mathbf{e}_2 = \frac{1}{\sqrt{2} |\varphi_0|} 
\left(
\begin{array}{c}
\langle \varphi \rangle \\
- \langle \bar{\varphi} \rangle \\
0 \\
0
\end{array}
\right), 
\notag \\
&
t_1 > 0 \ : \ \mathbf{e}_3 = \frac{1}{\sqrt{2} |\varphi_0|} 
\left(
\begin{array}{c}
0 \\
0 \\
- \langle \varphi \rangle \\
\langle \bar{\varphi} \rangle
\end{array}
\right), 
\qquad 
t_2 = 0 \ : \ \mathbf{e}_4 = \frac{1}{\sqrt{2} |\varphi_0|} 
\left(
\begin{array}{c}
0 \\
0 \\
\langle \varphi \rangle \\
\langle \bar{\varphi} \rangle
\end{array}
\right).
\end{align}
The unitary matrices are 
\begin{align}
& 
U_1 = 
\frac{1}{\sqrt{2} |\varphi_0|}
\left(
\begin{array}{cc}
\langle \bar{\varphi} \rangle & \langle \varphi \rangle \\
- \langle \bar{\varphi} \rangle & \langle \varphi \rangle 
\end{array}
\right), \qquad 
U_2 = 
\frac{1}{\sqrt{2} |\varphi_0|}
\left(
\begin{array}{cc}
- \langle \bar{\varphi} \rangle & \langle \varphi \rangle \\
 \langle \bar{\varphi} \rangle & \langle \varphi \rangle 
\end{array}
\right).
\label{eq:um}
\end{align}

We are faced with the fact that there are modes whose quadratic kinetic
terms disappear in the 
transverse
 ($s_2 = 0$) and the modulation
($t_2 = 0$) sectors.
In order to understand the nature of the zero eigenvalues of
$\mathbf{M}$, we analyze the broken generator of the symmetry in the vacuum.
The vacuum vector $\vec{v}$ is non-zero only in the modulation sector. 
Namely we have $\langle \partial_{\hat{m}} \varphi \rangle = 0$ and 
$\langle \partial_1 \varphi \rangle = i p \varphi_0 e^{ipx^1} = v$.
Therefore 
\begin{align}
\vec{v} = 
\left(
\begin{array}{c}
0 \\
0 \\
i p \varphi_0 e^{i p x^1} \\
- i p \bar{\varphi}_0 e^{-ipx^1}
\end{array}
\right).
\end{align}
The action of the translation $P^1$ and the $U(1)$ transformation
generators $T_{P^1}, T_{U(1)}$ on the VEVs are given by 
\begin{align}
& 
T_{P^1} v = i p \varphi_0 p e^{ipx^1}, \quad 
T_{P^1} \bar{v} = - i p \bar{\varphi}_0 p e^{-ipx^1},
\notag \\
&
T_{U(1)} v = i p \varphi_0 p \varphi_0 e^{ip x^1}, \quad
T_{U(1)} \bar{v} = - i p \bar{\varphi}_0 p e^{-ipx^1}.
\end{align}
The generator associated with the unbroken symmetry is defined
by $T_{\text{ub}} = T_{P^1} - T_{U(1)}$.
Indeed, we find 
that the action of $T_{\text{ub}}$ on $\vec{v}$ gives the vanishing result
$T_{\text{ub}} \vec{v} = 0$.
On the other hand, the generator associated with the broken symmetry
 is given by $T_{\text{b}} = T_{P^1} + T_{U(1)}$.
Then we find 
\begin{align}
T_{\text{b}} \vec{v} =
\left(
\begin{array}{c}
0 \\
0 \\
T_{\text{b}} v \\
- T_{\text{b}} \bar{v} 
\end{array}
\right)
= 
\left(
\begin{array}{c}
0 \\
0 \\
i p (p + p) \varphi_0 e^{ipx^1} \\
i p (p + p) \bar{\varphi}_0 e^{-ipx^1}
\end{array}
\right) = 
2 i p^2
\left(
\begin{array}{c}
0 \\
0 \\
\langle \varphi \rangle \\
\langle \bar{\varphi} \rangle 
\end{array}
\right)
\propto 
\mathbf{e}_4.
\end{align}
This is exactly the eigenvector for the zero eigenvalue $t_2 = 0$ in the
modulation sector. We note that the other zero eigenvalue $s_2=0$
corresponds to the flat direction in the $SO(2,1)$ invariant sector
which does not accompany with the spontaneous symmetry breaking.

By using the unitary matrix $U_2$ in \eqref{eq:um}, 
the matrix $M_2$ is diagonalized and 
we derive the Higgs and the NG modes associated with $t_1 > 0, t_2 = 0$ as 
\begin{align}
\tilde{\varphi}_{\text{H}} =
\frac{1}{\sqrt{2} |\varphi_0|}
\left(
- \langle \bar{\varphi} \rangle \partial_{1} \tilde{\varphi} +
 \langle \varphi \rangle \partial_{1} \tilde{\varphi}^{\dagger}
\right), \qquad 
\tilde{\varphi}_{\text{NG}} =
\frac{1}{\sqrt{2} |\varphi_0|}
\left(
\langle \bar{\varphi} \rangle \partial_{1} \tilde{\varphi} +
 \langle \varphi \rangle \partial_{1} \tilde{\varphi}^{\dagger}
\right).
\label{eq:NGHG}
\end{align}
These are the linear combinations of the fields that contain the
derivative along the $x^1$-direction.
It is natural to define the modes \eqref{eq:NGHG} as the derivative of the Higgs and NG fields
$\partial_1 H (x)$ and $\partial_1 N (x)$.
As we have clarified, the quadratic term of the NG field $N (x)$
vanishes in the energy functional (Lagrangian) while the Higgs field 
$H (x)$ appears without a mass gap.

Although, the NG and the Higgs modes emerge as an
derivative modes in the modulation direction, we 
are interested in the modes that propagate in the transverse
directions to the modulation.
In order to examine this, we perform the linear transformation of the upper half part 
of $\vec{\varphi}$ and diagonalize the left upper half of $\mathbf{M}$.
Using the unitary matrix $U_1$ in \eqref{eq:um}, the matrix $M_1$ is
diagonalized and we find that the Lagrangian in the quadratic order of fields is rewritten as 
\begin{align}
\mathcal{L}_{\text{quad.}} = - \frac{1}{2}
\left(
\tilde{\varphi}^{\dagger}_{\hat{m},c}, \tilde{\varphi}_{\hat{m},0},
 \tilde{\varphi}_{\text{H}}, \tilde{\varphi}_{\text{NG}} 
\right)
\left(
\begin{array}{cccc}
s_1 & 0 & 0 & 0 \\
0 & s_2 & 0 & 0 \\
0 & 0 & t_1 & 0 \\
0 & 0 & 0 & t_2
\end{array}
\right)
\left(
\begin{array}{c}
\tilde{\varphi}^{\hat{m}}_{c} \\
\tilde{\varphi}^{\dagger \hat{m}}_0 \\
\tilde{\varphi}_{\text{H}} \\
\tilde{\varphi}_{\text{NG}}^{\dagger}
\end{array}
\right).
\end{align}
Here and hereafter, 
the index $\hat{m}$ is contracted by $\eta_{\hat{m} \hat{n}} =
\text{diag.} (-1,1,1)$, and we have defined 
\begin{align}
& 
\tilde{\varphi}_{\hat{m}, \text{c}} = 
\frac{1}{\sqrt{2} |\varphi_0|}
\left(
\langle \bar{\varphi} \rangle \partial_{\hat{m}} \tilde{\varphi} +
 \langle \varphi \rangle \partial_{\hat{m}} \tilde{\varphi}^{\dagger}
\right), \qquad 
\tilde{\varphi}_{\hat{m}, 0}  =
\frac{1}{\sqrt{2} |\varphi_0|}
\left(
- \langle \bar{\varphi} \rangle \partial_{\hat{m}} \tilde{\varphi} +
 \langle \varphi \rangle \partial_{\hat{m}} \tilde{\varphi}^{\dagger}
\right).
\label{eq:c0}
\end{align}
Here $\tilde{\varphi}_{\hat{m},c}$ is the 
mode associated with $s_1 > 0$
and therefore it has the quadratic canonical kinetic term.
On the other hand, $\tilde{\varphi}_{\hat{m},0}$ is
the one for 
 $s_2 = 0$ whose quadratic kinetic term vanishes.
Note that they are distinguished from the Higgs and the NG modes in the
modulation direction.

Again we define the fields $\partial_{\hat{m}} A$, 
$\partial_{\hat{m}} B$ that correspond to the modes \eqref{eq:c0}. 
Then, 
the linear transformations \eqref{eq:c0} are interpreted as the following
field re-definition:\footnote{
Finding a field re-definition that realizes $H(x), N(x)$ in terms of
$\tilde{\varphi}$ is not straightforward since the VEV 
$\langle \varphi \rangle$ is not
a constant in the modulated vacuum. One finds that removing the
derivative $\partial_{\hat{m}}$ by an integration
in \eqref{eq:c0} is trivial but in \eqref{eq:NGHG} it is not.
}
\begin{align}
A (x) = \frac{1}{\sqrt{2} |\varphi_0|} 
\left(
\langle \bar{\varphi} \rangle \tilde{\varphi} + \langle \varphi \rangle \tilde{\varphi}^{\dagger}
\right), \qquad 
B = \frac{1}{\sqrt{2} |\varphi_0|} 
\left(
- \langle \bar{\varphi} \rangle \tilde{\varphi} + \langle \varphi
 \rangle \tilde{\varphi}^{\dagger}
\right).
\label{eq:AB}
\end{align}
Then, 
the NG and the Higgs modes in the modulation directions are
represented in terms of $A,B$:
\begin{align}
\tilde{\varphi}_{\text{NG}} = \partial_1 A - i p B, 
\qquad 
\tilde{\varphi}_{\text{H}} = \partial_1 B - i p A.
\label{eq:NG_Higgs_AB}
\end{align}
We obtain the Lagrangian 
for the dynamical fields $A,B$ in the modulated vacuum 
at the quadratic order as 
\begin{align}
\mathcal{L}_{\text{quad.}} =& \
- \frac{1}{2} s_1 \partial_{\hat{m}} A \partial^{\hat{m}} A
- \frac{1}{2} t_1 |\partial_1 B - i p A|^2.
\end{align}
One observes that the field $A$ does not propagate in the modulation
direction while $B$ does not propagate in the transverse direction.
Only the gradient of $B$ in the modulation direction contributes to the energy.
This is a reflection of the fact that the term $\partial_1 A$ is included in
the NG mode $\tilde{\varphi}_{\text{NG}}$ and it never appears in the
Lagrangian at the quadratic order.
A similar analysis has been done in \cite{Lauscher:2000ux}
 for the dispersion relation of NG and Higgs modes in a plane-wave type 
ground state in a Lorentz non-invariant theory.

We did not consider a potential term for $\varphi$, and consequently 
the system has the shift symmetry in Eq.~(\ref{eq:shift}).
What we have identified as a "Higgs boson" here is actually 
an NG boson associated with the spontaneous breaking of the shift symmetry.
If we add a potential term in the original Lagrangian, 
the Higgs boson obtains a mass. 
Therefore, the gapless property is originated from the shift symmetry, 
but what we have found here is the existence of the quadratic kinetic term 
of the Higgs boson, in contrast to its absence for the NG boson.

\subsection{Higher order terms}

Here, we study higher order expansion, and show that 
the cubic order of the expansion of 
 the Lagrangian
contains no term consisting of only the NG boson 
$\tilde\varphi_\text{NG}^3$, 
while in the quartic order 
there exists a term consisting of 
only the NG boson $\tilde\varphi_{\text{NG}}^4$. 
In general, we cannot exclude a possibility of 
$\tilde\varphi_{\text{NG}}^3$ 
a priori,
since the cubic derivative terms $(\der_1 \tilde\varphi)^3$ exist
after translational symmetry along $x^1$-direction 
 and the rotational symmetries in the $(x^1,x^2)$ and $(x^1,x^3)$
are broken.
To see this, we calculate the cubic 
derivative terms of fluctuation $\der \tilde\varphi$ 
in the Lagrangian \eqref{eq:Lagrangian}
with the introduction of 
the fluctuation $\varphi \to \vev{\varphi}+\tilde\varphi$
in \eqref{eq:fluct.}.
The explicit calculation leads to 
the cubic derivative terms of the Lagrangian 
${\cal L}_\text{cub.}$ as
\begin{equation}
\begin{split}
 {\cal L}_\text{cub.}
&=
ip
(2\lambda-9\alpha x_+)
(\vev{\varphi}\der_1 \tilde\varphi^\dg
-\vev{\bar\varphi} \der_1\tilde\varphi
)
|\der_1\tilde\varphi|^2
+
ip^3\alpha 
(
\vev{\varphi}^3(\der_1\tilde\varphi^\dg)^3
-\vev{\bar\varphi}^{ 3}(\der_1\tilde\varphi)^3)
\\
&\quad
+
ip(2\lambda -3\alpha x_+)\vev{\varphi}
(\der_1\tilde\varphi)(\der_{{\hat{m}}}\tilde\varphi^\dg)^2
-
ip(2\lambda -3\alpha x_+)\vev{\bar\varphi}
(\der_1\tilde\varphi^\dg)(\der_{{\hat{m}}}\tilde\varphi)^2
\\&\quad
-2ip\alpha x_+
(\vev{\varphi}\der_1 \tilde\varphi^\dg
-\vev{\bar\varphi} \der_1\tilde\varphi
)
|\der_{{\hat{m}}}\tilde\varphi|^2.
\end{split}
\end{equation}
By using Eq.~\eqref{eq:NGHG}, we find 
\begin{equation}
 {\cal L}_\text{cub.}
=i\sqrt{\frac{x_+}{2}}
(2\lambda +\alpha x_+ +4\alpha x_+)
\tilde\varphi_{\text{H}}\tilde\varphi_{\text{NG}}^2
+
i\sqrt{\frac{x_+}{2}}(-2\lambda +10\alpha x_+)\tilde\varphi_{\text{H}}^3
+\cdots,
\label{eq:cubNG}
\end{equation}
where the ellipsis $\cdots$ means that the terms with 
$\der_{\hat{m}} \tilde\varphi$. 
Thus, we find that the pure cubic NG term 
$\tilde\varphi_{\text{NG}}^3$
vanishes.

Now we see that 
the NG mode appears with a quartic derivative term.
The quartic derivative terms in the Lagrangian
\eqref{eq:Lagrangian} 
${\cal L}_{\text{quart.}}$ is similarly obtained as
\begin{equation}
\begin{split}
 {\cal L}_{\text{quart.}}
&=
(\lambda-9\alpha x_+)|\der_1\tilde\varphi|^4
+3\alpha p^2\vev{\varphi}^2
(\der_1\tilde\varphi^\dg)^2|\der_1\tilde\varphi|^2
+3\alpha p^2\vev{\bar\varphi}^{ 2}
(\der_1\tilde\varphi)^2|\der_1\tilde\varphi|^2
\\
&
\quad
+
(\lambda -3\alpha x_+)(\der_1\tilde\varphi)^2
(\der_{\hat{m}}\tilde\varphi^\dg)^2
+
(\lambda -3\alpha x_+)(\der_1\tilde\varphi^\dg)^2
(\der_{\hat{m}}\tilde\varphi)^2
\\
&
\quad
-3\alpha p^2\vev{\varphi}^2
 |\der_1\varphi|^2 (\der_{\hat{m}}\tilde\varphi^\dg)^2
-3\alpha p^2\vev{\bar\varphi}^{ 2}
 |\der_1\varphi|^2 (\der_{\hat{m}}\tilde\varphi)^2
\\
&
\quad
-4\alpha x_+ 
|\der_{\hat{p}}\tilde\varphi|^2
 |\der_1\tilde\varphi|^2
+\alpha p^2 \vev{\varphi}^2
|\der_{\hat{m}}\tilde\varphi|^2(\der_1 \tilde\varphi^\dg)^2
+\alpha p^2 \vev{\bar\varphi}^{ 2}
|\der_{\hat{m}}\tilde\varphi|^2(\der_1 \tilde\varphi)^2
\\
&
\quad
+
(\lambda-\alpha x_+)
(\der_{\hat{m}}\tilde\varphi)^2
(\der_{\hat{n}}\tilde\varphi^\dg)^2
+\alpha p^2\vev{\varphi}^2
|\der_{\hat{m}}\tilde\varphi|^2 (\der_{\hat{n}}\tilde\varphi^\dg)^2
+\alpha p^2 \vev{\bar\varphi}^{ 2}
|\der_{\hat{m}}\tilde\varphi|^2
(\der_{\hat{n}}\tilde\varphi)^2
.
\end{split}
\end{equation}
The quartic derivative term containing purely
NG mode $\tilde\varphi_{\text{NG}}$
is found by using \eqref{eq:NGHG} as
\begin{equation}
{\cal L}_{\text{quart.}}
=\frac{1}{4}(\lambda -6\alpha x_+)\tilde\varphi_{\text{NG}}^4+\cdots, 
\end{equation}
where the ellipsis $\cdots$ expresses
the terms other than $\tilde\varphi_{\text{NG}}^4$.
Therefore, we conclude that 
a term consisting of only the NG mode
appears with the 
quartic derivative term.

\section{Conclusion and discussions}
In this paper, we have studied the spatially modulated vacua in
a Lorentz invariant field theory where no finite density/temperature effects are included.
The NG theorem, for a global symmetry spontaneously broken due to 
vacuum expectation values of 
space-time derivatives of fields, 
states that there appears an NG boson 
without a canonical quadratic kinetic term but with 
a quadratic derivative term in the modulated direction
and a Higgs boson.  
We demonstrated this in the simple model 
whose energy functional can be written by the derivative
terms of the scalar fields.
The potential for the derivative terms allows
a local vacuum as the modulated vacuum, where 
translational symmetry along one direction which 
we choose $x^1$ and 
the rotational symmetries in the $(x^1,x^2), (x^1,x^3)$ planes
are spontaneously broken 
together with the $U(1)$ symmetry. 
A simultaneous transformation of $P^1$ and $U(1)$ is preserved in the
modulated vacuum. 
This modulated vacuum is completely consistent with
the equation of motion.
The vacuum energy depends on the parameters. 
In this paper, we have focused on the vacuum where the modulation takes
place only in the $x^1$-direction. 
Since the transverse directions $x^{\hat{m}}, (\hat{m} = 0,2,3)$ preserve the
$SO(2,1)$ Lorentz symmetry, there are no mixing terms between
$\partial_1 \varphi$ and $\partial_{\hat{m}} \varphi$ in the Lagrangian in the quadratic
order of fields. This results in the complete block diagonal form of the
matrix $\mathbf{M}$ in \eqref{eq:vector_matrix}.
We have explicitly shown that the ``mass eigenstates'' in the modulation
and the transverse directions are different.
Therefore we are not able to perform the diagonalization in these
directions simultaneously.
We have employed the linear combination of the fields \eqref{eq:AB} and
represented the NG and Higgs modes in terms of $A$ and $B$. 
The $A$-mode propagates in the transverse directions while the $B$-mode
only oscillates in the modulation direction.
Finally, we have demonstrated that 
a term containing only NG modes appears in the quartic 
derivative order.
The Higgs mode, which is defined as the orthogonal mode to the NG mode
in our discussion, is indeed
an NG mode associated with the spontaneously broken shift symmetry.
We note that this Higgs mode has its non-zero quadratic kinetic term.
This is a consequence of the application of the NG theorem to higher
derivative field theories.

Although we have illustrated the stability of the modulated vacuum 
in our simplest model, 
we would like to emphasize that 
the stability analysis employed in this paper
is general at the linear level 
for any model admitting local vacuum exhibiting 
a spatial modulation.

Among general solutions in Eq.~(\ref{eq:general_sol}) which are energetically degenerated, we have focused on the FF state, which has the highest unbroken symmetry. 
Which vacuum is chosen among energetically degenerated vacua with
different unbroken symmetry is known as a vacuum alignment problem first
discussed in the context of technicolor models
\cite{Peskin:1980gc,Preskill:1980mz}. In such the cases, quantum
corrections pick up the vacuum with the highest unbroken symmetry, and
therefore we expect the same happens in our case too.
We note that the structure of the vacuum modulation crucially depends on models we consider.
For example, inhomogeneous chiral condensates in dense QCD appears to be a FFLO instead of a FF one.

We have studied the modulated vacuum in
the Ginzburg-Landau type effective theory
in the Lorentz invariant framework,
without assuming any underlying microscopic theory.
However, there is an argument about a no-go theorem
for modulated vacua by fermion condensates
in relativistic QCD-like theories \cite{Splittorff:2000mm}.
It is an interesting open question whether
our model can be obtained as the low-energy theory
of a fermion condensation in relativistic theories or not.
Possible future directions are in order.
Beyond the semiclassical level in this paper,
we need a more rigorous proof for the generalized NG theorem in full quantum level. 
The Higgs mechanism in a $U(1)$-gauged model, 
and spatial modulations along two or more directions \cite{Kojo:2011cn}
as well as a temporal modulation \cite{Hayata:2013sea} 
are interesting directions.
Applying our discussion to 
more general higher derivative theories 
such as higher-order Skyrme model \cite{Gudnason:2017opo} 
is also one 
of future directions.
We also would like to
embed our model to 
supersymmetric theories based on the formalism in 
Ref.~\cite{Nitta:2014fca}, 
and supersymmetry breaking in modulated vacua will be reported elsewhere 
\cite{Nitta:2017yuf}.

\subsection*{Acknowledgments}

The authors would like to thank N.~Yamamoto for notifying the reference
\cite{Splittorff:2000mm} and the exclusion of the FFLO phase in relativistic QCD-like theories.
The work of M.~N.~is 
supported in part by 
the Ministry of Education,
Culture, Sports, Science (MEXT)-Supported Program for the Strategic Research Foundation at Private Universities `Topological Science' (Grant No.\ S1511006), 
the Japan Society for the Promotion of Science
(JSPS) Grant-in-Aid for Scientific Research (KAKENHI Grant
No.~16H03984), 
and a Grant-in-Aid for
Scientific Research on Innovative Areas ``Topological Materials
Science'' (KAKENHI Grant No.~15H05855) from the MEXT of Japan. 
The work of S.~S. is supported in part by JSPS KAKENHI Grant Number
JP17K14294 and Kitasato University Research Grant for Young Researchers.
The work of R.~Y.~is 
supported by Research Fellowships of JSPS
 for Young Scientists Grant Number 16J03226.

\end{document}